\documentclass[aps,preprint]{revtex4}%
\usepackage{amsfonts}
\usepackage{amsmath}
\usepackage{amssymb}
\usepackage{graphicx}%
\begin{document}
\preprint{CTP-SCU/2015022}
\title{The Classical Limit of Minimal Length Uncertainty Relation: Revisit with the
Hamilton-Jacobi Method}
\author{Xiaobo Guo$^{a}$}
\email{guoxiaobo@swust.edu.cn}
\author{Peng Wang$^{b}$}
\email{pengw@scu.edu.cn}
\author{Haitang Yang$^{b,c}$}
\email{hyanga@scu.edu.cn}
\affiliation{$^{a}$School of Science, Southwest University of Science and Technology,
Mianyang, 621010, PR China }
\affiliation{$^{b}$Center for Theoretical Physics, College of Physical Science and
Technology, Sichuan University, Chengdu, 610064, PR China}
\affiliation{$^{c}$Kavli Institute for Theoretical Physics China (KITPC), Chinese Academy
of Sciences, Beijing 100080, P.R. China}

\begin{abstract}
The existence of a minimum measurable length could deform not only the
standard quantum mechanics but also classical physics. The effects of the
minimal length on classical orbits of particles in a gravitation field have
been investigated before, using the deformed Poisson bracket or Schwarzschild
metric. In this paper, we use the Hamilton-Jacobi method to study motions of
particles in the context of deformed Newtonian mechanics and general
relativity. Specifically, the precession of planetary orbits, deflection of
light, and time delay in radar propagation are considered in this paper. We
also set limits on the deformation parameter by comparing our results with the
observational measurements. Finally, comparison with results from previous
papers is given at the end of this paper.

\end{abstract}
\maketitle
\tableofcontents



\section{Introduction}

One of the predictions shared by various quantum theories of gravity is the
existence of a minimal observable length. For example, this fundamental
minimal length scale could arise in the framework of the\ string theory
\cite{IN-Veneziano:1986zf,IN-Gross:1987ar,IN-Amati:1988tn}. For a review of a
minimal length in quantum gravity, see \cite{IN-Garay:1994en}. Some
realizations of the minimal length from various scenarios have been proposed.
Specifically, one of popular models is the generalized uncertainty principle
(GUP) \cite{IN-Maggiore:1993kv,IN-Kempf:1994su} derived from the deformed
fundamental commutation relation:
\begin{equation}
\lbrack X,P]=i\hbar(1+\beta P^{2}), \label{eq:1dGUP}%
\end{equation}
where $\beta$ is some deformation parameter, and the minimal measurable length
is $\Delta_{\text{min}}=\hbar\sqrt{\beta}$. For a review of the GUP, see
\cite{IN-Hossenfelder:2012jw}. The deformed fundamental commutator $\left(
\ref{eq:1dGUP}\right)  $ have been widely discussed in the context of quantum
mechanics, such as the harmonic oscillator \cite{IN-Chang:2001kn}, Coulomb
potential\cite{IN-Akhoury:2003kc,IN-Brau:1999uv}, and gravitational well
\cite{IN-Brau:2006ca,IN-Pedram:2011xj}. Since there is a UV-IR mixing embodied
in the deformed commutation relation \cite{IN-Benczik:2002tt}, it is also
important to study effects of the minimal length in a classical context. For
example, the effects of GUP on the classical quantum cosmology were discussed
in \cite{IN-Jalalzadeh:2014jea}, effects on classical harmonic oscillator in
\cite{IN-Quintela:2015bua}, and effects on equivalence principle in
\cite{IN-Tkachuk:2013qa}.

General relativity is the standard theory of gravity. The observational tests
of gravity have been performed on Earth and in the solar system, such as the
procession of the perihelia of orbit of Mercury, deflection of light by the
Sun, and time delay of radar echoes passing the Sun. To set limits on new
physics beyond General relativity, effects of the deformed commutation
relation on these observational tests have been considered in
\cite{IN-Benczik:2002tt,IN-Ahmadi:2014cga,IN-Silagadze:2009vu,IN-Scardigli:2014qka,IN-Ali:2015zua}%
. Specifically, by replacing the deformed quantum mechanical commutator by the
deformed\ Poisson bracket via%
\begin{equation}
\frac{1}{i\hbar}\left[  \hat{A},\hat{B}\right]  \Rightarrow\left\{
A,B\right\}  ,
\end{equation}
the authors of \cite{IN-Benczik:2002tt,IN-Silagadze:2009vu,IN-Ahmadi:2014cga}
found equations of motion and orbit of Mercury in the context of deformed
Newtonian dynamics. Since the procession of the perihelia of orbit of Mercury
is predicted by general relativity not Newtonian mechanics, it is more
appropriate to study it in the context of deformed general relativity.
Furthermore, it is impossible to calculate the trajectory of a photo in
deformed Newtonian dynamics. Motivated by these considerations, the authors of
\cite{IN-Scardigli:2014qka} proposed a modification of the Schwarzschild
metric to reproduce the modified Hawking temperature derived from the deformed
fundamental commutation relation $\left(  \ref{eq:1dGUP}\right)  $. Using this
deformed metric, they computed corrections to the standard general
relativistic predictions for the light deflection and perihelion precession.
In \cite{IN-Scardigli:2014qka}, only the metric was deformed and the equation
of motion of a test particle was still given by the standard geodesic
equation. As pointed out in \cite{IN-Scardigli:2014qka}, a more profound way
to obtain the geodesic in deformed general relativity would be from the
deformed field equations of general relativity not just assuming a deformed
solution (as in \cite{IN-Scardigli:2014qka}), or a deformed kinematics (as in
\cite{IN-Benczik:2002tt}). However, such deformed field equations are not
available yet. Alternatively, the geodesics can be obtained using the
Hamilton-Jacobi method. In \cite{IN-Tao:2012fp}, we discussed the
Hamilton-Jacobi method in the context of deformed $1$D Newtonian mechanics.
Moreover, the deformed Hamilton-Jacobi equations in curved spacetime have been
derived when corrections, caused by the deformed fundamental commutator
$\left(  \ref{eq:1dGUP}\right)  $, to the Hawking temperature were studied
using the Hamilton-Jacobi method
\cite{IN-Chen:2013tha,IN-Chen:2013ssa,IN-Chen:2014xgj}.

In this paper, we use the Hamilton-Jacobi method to study effects of the
minimal length on geodesic motions of particles in the context of deformed
Newtonian dynamics and general relativity. Concretely, in section \ref{Sec:NR}
we calculate the precession angle of planetary orbits in the context of
deformed Newtonian dynamics after the deformed Hamilton-Jacobi equation is
derived. It turns out that our result $\left(  \ref{eq:procession-NM}\right)
$ agrees with these obtained in \cite{IN-Benczik:2002tt} with $\beta^{\prime
}=2\beta$ and \cite{IN-Silagadze:2009vu}, where the method of the
deformed\ Poisson bracket was used. In section \ref{Sec:RC}, we derive the
deformed Hamilton-Jacobi equations in curved spacetime and the precession
angle of planetary orbits in the context of deformed general relativity.
Contrary to what was found in \cite{IN-Scardigli:2014qka}, our results show
that the leading correction to the precession angle caused by deformations
coincides with these obtained in \cite{IN-Benczik:2002tt,IN-Silagadze:2009vu}.
This discrepancy may come from an implicit assumption made in
\cite{IN-Scardigli:2014qka} about the energy of planets, which is discussed in
detail in section \ref{Sec:Con}. The deflection of light and time delay in
radar propagation are also considered in section \ref{Sec:RC}. We place
constraints on the deformation parameter by comparing our results with the
observational measurements in section \ref{Sec:CE}. Section \ref{Sec:Con} is
devoted to our discussion and conclusion. For simplicity, we set $c=k_{B}=1$
in this paper.

\section{Hamilton-Jacobi Method in Deformed Newtonian Dynamics}

\label{Sec:NR}

In this section, we first derive the deformed Hamilton-Jacobi equation for a
nonrelativistic system and then apply it to the motion in a central potential.

\subsection{Deformed Hamilton-Jacobi Equation}

In three dimensions, a generalization of the deformed algebra $\left(
\ref{eq:1dGUP}\right)  $ reads \cite{IN-Kempf:1994su}
\begin{align}
\lbrack X_{i},P_{j}]  &  =i\hbar\left[  (1+\beta P^{2})\delta_{ij}%
+\beta^{\prime}P_{i}P_{j}\right]  \text{,}\nonumber\\
\left[  X_{i},X_{j}\right]   &  =i\hbar\frac{\left(  2\beta-\beta^{\prime
}\right)  +\left(  2\beta+\beta^{\prime}\right)  \beta P^{2}}{1+\beta P^{2}%
}\left(  P_{i}X_{j}-P_{j}X_{i}\right)  \text{,}\label{eq:deformedC}\\
\left[  P_{i},P_{j}\right]   &  =0\text{,}\nonumber
\end{align}
where $\beta$, $\beta^{\prime}>0$ are two deformation parameters, and the
minimal length becomes $\Delta X_{\min}=\hbar\sqrt{\beta+\beta^{\prime}}$. To
study the Schrodinger equation incorporating the minimal length commutation
relations $\left(  \ref{eq:deformedC}\right)  $, we need the representations
of $X_{i}$ and $P_{i}$ in terms of some differential operators. In our paper,
we consider the Brau reduction \cite{IN-Brau:1999uv}, where $\beta^{\prime
}=2\beta$ and the commutators taken between different components of the
position $X_{i}$ vanish to first order in $\beta$ and $\beta^{\prime}$.

For this particular case, there is a very simple reduction of the form to
first order in $\beta$:%
\begin{align}
X_{i}  &  =x_{i},\nonumber\\
P_{i}  &  =p_{i}\left(  1+\beta p^{2}\right)  , \label{eq:xandp}%
\end{align}
where $x_{i}$ and $p_{i}$ are the conventional momentum and position operators
satisfying
\begin{equation}
\left[  x_{i},p_{j}\right]  =i\hbar\delta_{ij}\text{, }\left[  x_{i}%
,x_{j}\right]  =\left[  p_{i},p_{j}\right]  =0\text{,}%
\end{equation}
and $p^{2}=%
{\displaystyle\sum\limits_{i}}
p_{i}p_{i}$. In the pseudo-position representation, one then has%
\begin{equation}
x_{i}=x_{i}\text{ and }p_{i}=\frac{\hbar}{i}\frac{\partial}{\partial x_{i}}.
\end{equation}
For simplicity, we could put eqns. $\left(  \ref{eq:xandp}\right)  $ in a
form:
\begin{align}
X_{i}  &  =x_{i},\nonumber\\
P_{i}  &  =p_{i}f\left(  \beta p^{2}\right)  ,
\end{align}
where $f\left(  x\right)  =1+x$ for the Brau reduction.

We now consider a nonrelativistic quantum system with a Hamiltonian of the
general form:%
\begin{equation}
H=\frac{\vec{P}\cdot\vec{P}}{2m}+V\left(  X\right)  .
\end{equation}
The deformed time dependent Schrodinger equation is%
\begin{equation}
\frac{\vec{P}\cdot\vec{P}}{2m}\psi\left(  \vec{x},t\right)  +V\left(
x\right)  \psi\left(  \vec{x},t\right)  =i\hbar\frac{\partial\psi\left(
\vec{x},t\right)  }{\partial t}, \label{eq:TDSE}%
\end{equation}
where $\vec{P}=\vec{p}f\left(  \beta p^{2}\right)  $. Substituting the ansatz
$\psi\left(  x,t\right)  =\exp\left[  \frac{iS\left(  \vec{x},t\right)
}{\hbar}\right]  $ into eqn. $\left(  \ref{eq:TDSE}\right)  $ and taking the
limit $\hbar\rightarrow0$, one finds that the leading order of eqn. $\left(
\ref{eq:TDSE}\right)  $ gives the classical Hamilton-Jacobi equation in
deformed spaces:%
\begin{equation}
\frac{1}{2m}\left(  \vec{\nabla}S\cdot\vec{\nabla}S\right)  f^{2}\left(
\beta\vec{\nabla}S\cdot\vec{\nabla}S\right)  +V\left(  \vec{x}\right)
+\frac{\partial S}{\partial t}=0, \label{eq:DHJE}%
\end{equation}
where $S\left(  \vec{x},t\right)  $ is the classical action, and $\vec{p}%
\psi=\vec{\nabla}S\psi$. If the potential $V\left(  \vec{x}\right)  $ does not
depend explicitly on time, we can separate the variables as%
\begin{equation}
S=W\left(  \vec{x}\right)  -Et,
\end{equation}
where $E$ can be identified with the total energy.

\subsection{Motion in a Central Potential}

When a particle is moving in a central potential $V\left(  r\right)  $, the
Hamilton-Jacobi equation can be solved in the spherical coordinates. In the
spherical coordinates, one has for $\vec{\nabla}$%
\begin{equation}
\vec{\nabla}=\hat{r}\frac{\partial}{\partial r}+\frac{\hat{\theta}}{r}%
\frac{\partial}{\partial\theta}+\frac{\hat{\phi}}{r\sin\theta}\frac{\partial
}{\partial\phi}.
\end{equation}
Thus, we find%
\begin{equation}
\vec{\nabla}S\cdot\vec{\nabla}S=\left(  \frac{\partial S}{\partial r}\right)
^{2}+\frac{1}{r^{2}}\left(  \frac{\partial S}{\partial\theta}\right)
^{2}+\frac{1}{r^{2}\sin^{2}\theta}\left(  \frac{\partial S}{\partial\phi
}\right)  ^{2}.
\end{equation}
Since there are no explicit $t$- and $\phi$-dependence in the Hamilton-Jacobi
equation, we assume that%
\begin{equation}
S=S_{1}\left(  r\right)  +S_{2}\left(  \theta\right)  -Et+L_{z}\phi,
\end{equation}
where $E$ and $L_{z}$ have the meaning of the energy and $z$-component of the
orbital angular momentum, respectively. To separate the variable $\theta$ from
$r$, one can introduce a constant $L$ and has the equation for $S_{2}\left(
\theta\right)  $
\begin{equation}
\left(  \frac{dS_{2}}{d\theta}\right)  ^{2}+\frac{L_{z}^{2}}{\sin^{2}\theta
}=L^{2},
\end{equation}
where $L$ represents the orbital angular momentum. The equation for
$S_{1}\left(  r\right)  $ then becomes%
\begin{equation}
\frac{1}{2m}\left[  \left(  \frac{dS_{1}}{dr}\right)  ^{2}+\frac{L^{2}}{r^{2}%
}\right]  f^{2}\left(  \beta\left[  \left(  \frac{dS_{1}}{dr}\right)
^{2}+\frac{L^{2}}{r^{2}}\right]  \right)  +V\left(  r\right)  =E.
\label{eq:S1}%
\end{equation}
For the function $f\left(  x\right)  $, one can solve the equation
$y=xf^{2}\left(  x\right)  $ for $x$ in terms of $y$ and express the solution
as $x=yg\left(  y\right)  $. For the Brau reduction, we have $g\left(
x\right)  =1-2x+\mathcal{O}\left(  x^{2}\right)  $. Solving eqn. $\left(
\ref{eq:S1}\right)  $ gives%
\begin{equation}
S_{1}=\int dr\sqrt{2m\left[  E-V\left(  r\right)  \right]  g\left(
2m\beta\left[  E-V\left(  r\right)  \right]  \right)  -\frac{L^{2}}{r^{2}}}.
\end{equation}
We can choose the $z$-axis such that the motion of the particle is in the
$x$-$y$ plane. Then $\sin\theta=1$ and $L=L_{z}$. In this case, we use the
inverse Legendre transform to find \cite{NC-Goldstein}%
\begin{equation}
\phi=-\frac{\partial S_{1}}{\partial L}=L\int\frac{dr}{r^{2}\sqrt{2m\left[
E-V\left(  r\right)  \right]  g\left(  2m\beta\left[  E-V\left(  r\right)
\right]  \right)  -\frac{L^{2}}{r^{2}}}}. \label{eq:phiNR}%
\end{equation}

Considering the Kepler motion with $E<0$, we have a specific form of the
potential:%
\begin{equation}
V\left(  r\right)  =-\frac{k}{r},
\end{equation}
where $k=GMm$. In this case, eqn. $\left(  \ref{eq:phiNR}\right)  $ takes on
the form:%
\begin{equation}
\phi=\int\frac{d\tilde{r}}{\tilde{r}}\frac{1-\varepsilon}{\sqrt{\left(
\tilde{r}-\tilde{r}_{\min}\right)  \left(  \tilde{r}_{\max}-\tilde{r}\right)
}}+\mathcal{O}\left(  \varepsilon^{2}\right)  , \label{eq:phiNR2}%
\end{equation}
where
\begin{equation}
\varepsilon=2m\beta\left\vert E\right\vert ,\text{ }A=\frac{k}{L}\sqrt
{\frac{2m}{\left\vert E\right\vert }},\text{ }\tilde{r}=\frac{2m\left\vert
E\right\vert r}{L^{2}},
\end{equation}
and
\begin{equation}
\tilde{r}_{\max/\min}=\frac{A+4\varepsilon A\pm\sqrt{A^{2}-4-8\varepsilon}%
}{2+4\varepsilon}.
\end{equation}
It follows from eqn. $\left(  \ref{eq:phiNR2}\right)  $ that $\tilde{r}$
reached its minimum and maximum values $\tilde{r}_{\min}$ and $\tilde{r}%
_{\max}$ at perihelion and aphelion. Integrating eqn. $\left(  \ref{eq:phiNR2}%
\right)  $ leads to%
\begin{equation}
\phi\left(  \tilde{r}\right)  =\frac{\left(  1-\varepsilon\right)  }%
{\sqrt{\tilde{r}_{\min}\tilde{r}_{\max}}}\arccos\left(  \frac{2\tilde{r}%
_{\min}\tilde{r}_{\max}-\left(  \tilde{r}_{\min}+\tilde{r}_{\max}\right)
\tilde{r}}{\tilde{r}\left(  \tilde{r}_{\max}-\tilde{r}_{\min}\right)
}\right)  -\phi_{0}+\mathcal{O}\left(  \varepsilon^{2}\right)  ,
\label{eq:phi}%
\end{equation}
In particular, one finds from eqn. $\left(  \ref{eq:phi}\right)  $ that%
\begin{equation}
\phi\left(  \tilde{r}_{\max}\right)  -\phi\left(  \tilde{r}_{\min}\right)
=\left(  1-A^{2}\varepsilon\right)  \pi+\mathcal{O}\left(  \varepsilon
^{2}\right)  .
\end{equation}
It precesses by an angle of $-2\pi A^{2}\varepsilon$ per revolution. For
$\varepsilon\ll1$, the precession angle is%
\begin{equation}
\Delta\omega_{\beta}=-2\pi A^{2}\varepsilon+\mathcal{O}\left(  \varepsilon
^{2}\right)  . \label{eq:omega}%
\end{equation}
In the leading order, one has that%
\begin{equation}
e=\sqrt{1-\frac{2\left\vert E\right\vert L^{2}}{mk^{2}}}\text{ and }a=\frac
{k}{2\left\vert E\right\vert },
\end{equation}
where $a$ is the semi-major axis of the planet's orbit, and $e$ is it's
eccentricity. Thus, eqn. $\left(  \ref{eq:omega}\right)  $ becomes
\begin{equation}
\Delta\omega_{\beta}\approx-2\pi\frac{4\beta GMm^{2}}{a\left(  1-e^{2}\right)
}, \label{eq:procession-NM}%
\end{equation}
which perfectly coincides with the result of \cite{IN-Benczik:2002tt} (eqn.
$\left(  66\right)  $) with $\beta^{\prime}=2\beta$.

\section{Hamilton-Jacobi Method in Deformed Relativity Theory}

\label{Sec:RC}

In section \ref{Sec:NR}, we consider the nonrelativistic case and obtain the
deformed Hamilton-Jacobi equation $\left(  \ref{eq:DHJE}\right)  $ by using
WKB approximation to find the leading order in $\hbar$ of the deformed
Schrodinger equation $\left(  \ref{eq:TDSE}\right)  $. In this section, we
derive the deformed Hamilton-Jacobi equations in the relativistic case. To do
so, we first find the deformed Klein-Gordon, Dirac, and Maxwell's equations
incorporating the minimal length commutation relations $\left(
\ref{eq:deformedC}\right)  $. After the deformed Hamilton-Jacobi equations are
given, motions of massive and massless particles through the Schwarzschild
metric are investigated. We also discuss effects of the minimal length on the
experimental tests of general relativity, e.g. precession of planetary orbits,
the bending of light, and time-delay in radar propagation.

\subsection{Deformed Hamilton-Jacobi Equation}

When the deformed commutation relations $\left(  \ref{eq:deformedC}\right)  $
are considered, the deformed Klein-Gordon equation for a scalar particle of
mass $m$ has been suggested in \cite{RC-Kempf:1996nk,RC-Berger:2010pj}:
\begin{equation}
\left(  \partial_{t}^{2}+\frac{P^{2}}{\hbar^{2}}+\frac{m^{2}}{\hbar^{2}%
}\right)  \Phi=0,
\end{equation}
where
\begin{equation}
P_{i}=p_{i}\left(  1+\beta p^{2}\right)  \equiv p_{i}f\left(  \beta
p^{2}\right)  \text{, }p_{i}\Phi=\frac{\hbar}{i}\partial_{i}\Phi\text{,}%
\end{equation}
and the index $i$ runs over spatial coordinates. Expressing $\Phi$ in terms of
$S\left(  \vec{x},t\right)  $:%
\begin{equation}
\Phi\left(  x,t\right)  =\exp\left[  \frac{iS\left(  \vec{x},t\right)  }%
{\hbar}\right]  ,
\end{equation}
one finds the lowest order in $\hbar$ gives the deformed Hamilton-Jacobi
equation for a scalar particle:%
\begin{equation}
\left(  \frac{\partial S}{\partial t}\right)  ^{2}-\left(  \vec{\nabla}%
S\cdot\vec{\nabla}S\right)  f^{2}\left(  \beta\vec{\nabla}S\cdot\vec{\nabla
}S\right)  =m^{2}. \label{eq:HJFD}%
\end{equation}

Similarly, the deformed Dirac equation for a spin-$1/2$ fermion of mass $m$
takes the form:%
\begin{equation}
\left(  \gamma_{0}i\partial_{t}+\frac{\vec{\gamma}\cdot\vec{P}}{\hbar}%
-\frac{m}{\hbar}\right)  \Psi=0, \label{eq:DiracFD}%
\end{equation}
where $\gamma_{0}$ and $\vec{\gamma}$ are Gamma matrices. Multiplying $\left(
\gamma_{0}i\partial_{t}+\frac{\vec{\gamma}\cdot\vec{p}}{\hbar}+\frac{m}{\hbar
}\right)  $ by eqn. $\left(  \ref{eq:DiracFD}\right)  $ and using the gamma
matrices anticommutation relations, the deformed Dirac equation can be written
as%
\begin{equation}
\partial_{t}^{2}\Psi=-\left(  \frac{P^{2}}{\hbar^{2}}+\frac{m^{2}}{\hbar^{2}%
}\right)  \Psi+\frac{\left[  \gamma_{i},\gamma_{j}\right]  }{2}P_{i}P_{j}\Psi.
\label{eq:DiracFDM}%
\end{equation}
To obtain the Hamilton-Jacobi equation for the fermion, the ansatz for $\Psi$
takes the form:
\begin{equation}
\Psi=\exp\left[  \frac{iS\left(  \vec{x},t\right)  }{\hbar}\right]  v,
\label{eq:fermionansatzD}%
\end{equation}
where $v$ is a vector function of the spacetime. Substituting eqn. $\left(
\ref{eq:fermionansatzD}\right)  $ into eqn. $\left(  \ref{eq:DiracFDM}\right)
$ and noting that the second term on RHS of eqn. $\left(  \ref{eq:DiracFDM}%
\right)  $ does not contribute to the lowest order in $\hbar$, we thus find
that the deformed Hamilton-Jacobi equation for a fermion is the same as that
for a scalar, which is eqn. $\left(  \ref{eq:HJFD}\right)  $.

The deformed Maxwell's equations for a massless vector field $A_{\mu}$ is%
\begin{equation}
\tilde{\partial}^{\mu}\tilde{F}_{\mu\nu}=0, \label{eq:Maxwell-Equation}%
\end{equation}
where $\tilde{\partial}_{t}=\partial_{t}$, $\tilde{\partial}_{i}=f\left(
-\beta\hbar^{2}\partial_{i}\partial^{i}\right)  \partial_{i}$, and $\tilde
{F}_{\mu\nu}=\tilde{\nabla}_{\mu}A_{\nu}-\tilde{\nabla}_{\nu}A_{\mu}$. We then
make the WKB ansatz:%
\begin{equation}
A_{\mu}=a_{\mu}\exp\left(  \frac{iS\left(  \vec{x},t\right)  }{\hbar}\right)
,
\end{equation}
where $a_{\mu}$ is the polarization vector, and $S\left(  \vec{x},t\right)  $
is the action. Plugging the WKB ansatz into eqn. $\left(
\ref{eq:Maxwell-Equation}\right)  $, we find that leading order in $\hbar$
gives%
\begin{equation}
a_{\nu}S^{\nu}S_{\mu}-a_{\mu}S^{\nu}S_{\nu}=0, \label{eq:Hamilton-JacobiV}%
\end{equation}
where
\begin{equation}
S_{t}\equiv\frac{\partial S}{\partial t}\text{ and }S_{i}\equiv f\left(
\beta\vec{\nabla}S\cdot\vec{\nabla}S\right)  \partial_{i}S.
\end{equation}
To simplify eqn. $\left(  \ref{eq:Hamilton-JacobiV}\right)  $, one could
impose the Lorentz gauge:%
\begin{equation}
\tilde{\partial}^{\mu}A_{\mu}=0,
\end{equation}
whose leading order is
\begin{equation}
a_{\nu}S^{\nu}=0. \label{eq:Lorentz-Gauge}%
\end{equation}
By plugging eqn. $\left(  \ref{eq:Lorentz-Gauge}\right)  $ into eqn. $\left(
\ref{eq:Hamilton-JacobiV}\right)  $, it shows that the Hamilton-Jacobi
equation for a massless vector boson is also given by eqn. $\left(
\ref{eq:HJFD}\right)  $ with $m=0$.

\subsection{Motion in the Schwarzschild Metric}

We now generalize the deformed Hamilton-Jacobi equation $\left(
\ref{eq:HJFD}\right)  $ in flat spacetime to the Schwarzschild metric:%
\begin{equation}
ds^{2}=h\left(  r\right)  dt^{2}-\frac{dr^{2}}{h\left(  r\right)  }%
-r^{2}\left(  d\theta^{2}+\sin^{2}\theta d\phi^{2}\right)  ,
\label{eq:Schwarzschild metric}%
\end{equation}
where we have
\begin{equation}
h\left(  r\right)  =1-\frac{2GM}{r}.
\end{equation}
In \cite{IN-Chen:2014xgj}, the deformed Hamilton-Jacobi equations in the
Schwarzschild metric for scalars and spin $1/2$ fermions have been derived
from the deformed Klein-Gordon and Dirac equations in the Schwarzschild
metric. Here, we use an easier but less rigorous way to obtain the deformed
Hamilton-Jacobi equations in curved spacetime. First consider the
Hamilton-Jacobi equation without GUP modifications. In
\cite{RC-Benrong:2014woa}, we showed that the unmodified Hamilton-Jacobi
equation in curved spacetime with the metric $ds^{2}=g_{\mu\nu}dx^{\mu}%
dx^{\nu}$ was%
\begin{equation}
g^{\mu\nu}\partial_{\mu}S\partial_{\nu}S-m^{2}=0.
\end{equation}
Therefore, the unmodified Hamilton-Jacobi equation in the Schwarzschild metric
becomes
\begin{equation}
\frac{\left(  \partial_{t}S\right)  ^{2}}{h\left(  r\right)  }-h\left(
r\right)  \left(  \partial_{r}S\right)  ^{2}-\frac{\left(  \partial_{\theta
}S\right)  ^{2}}{r^{2}}-\frac{\left(  \partial_{\phi}S\right)  ^{2}}{r^{2}%
\sin^{2}\theta}=m^{2}. \label{eq:HJC}%
\end{equation}
On the other hand, the unmodified Hamilton-Jacobi equation in flat spacetime
can be obtained from eqn. $\left(  \ref{eq:HJFD}\right)  $ by taking $\beta
=0$:%
\begin{equation}
\left(  \partial_{t}S\right)  ^{2}-\left(  \partial_{r}S\right)  ^{2}%
-\frac{\left(  \partial_{\theta}S\right)  ^{2}}{r^{2}}-\frac{\left(
\partial_{\phi}S\right)  ^{2}}{r^{2}\sin^{2}\theta}=m^{2}. \label{eq:HJFF}%
\end{equation}
Comparing eqn. $\left(  \ref{eq:HJC}\right)  $ to eqn. $\left(  \ref{eq:HJFF}%
\right)  $, one finds that the Hamilton-Jacobi equation in the Schwarzschild
metric can be obtained from that in flat spacetime by making replacements
$\partial_{r}S\rightarrow\sqrt{h\left(  r\right)  }\partial_{r}S$ and
$\partial_{t}S\rightarrow\frac{\partial_{t}S}{\sqrt{h\left(  r\right)  }}$.
Likewise, by making replacements $\partial_{r}S\rightarrow\sqrt{h\left(
r\right)  }\partial_{r}S$ and $\partial_{t}S\rightarrow\frac{\partial_{t}%
S}{\sqrt{h\left(  r\right)  }}$, the deformed Hamilton-Jacobi equation in flat
spacetime $\left(  \ref{eq:HJFD}\right)  $ leads to that in the Schwarzschild
metric%
\begin{equation}
\frac{1}{h\left(  r\right)  }\left(  \frac{\partial S}{\partial t}\right)
^{2}-\mathcal{X}f^{2}\left(  \beta\mathcal{X}\right)  =m^{2}, \label{eq:DHJEE}%
\end{equation}
where we have
\begin{equation}
\mathcal{X}=h\left(  r\right)  \left(  \partial_{r}S\right)  ^{2}%
+\frac{\left(  \partial_{\theta}S\right)  ^{2}}{r^{2}}+\frac{\left(
\partial_{\phi}S\right)  ^{2}}{r^{2}\sin^{2}\theta}.
\end{equation}

Since there are no explicit $t$- and $\phi$-dependence in the Schwarzschild
metric, we assume that%
\begin{equation}
S=S_{1}\left(  r\right)  +S_{2}\left(  \theta\right)  -Et+L_{z}\phi,
\end{equation}
where $E$ and $L_{z}$ can be identified as the energy and $z$-component of the
orbital angular momentum, respectively. Introducing a constant $L$
representing the orbital angular momentum, one finds that $S_{2}\left(
\theta\right)  $ satisfies
\begin{equation}
\left(  \frac{dS_{2}}{d\theta}\right)  ^{2}+\frac{L_{z}^{2}}{\sin^{2}\theta
}=L^{2}.
\end{equation}
The equation for $S_{1}\left(  r\right)  $ is%
\begin{equation}
\mathcal{X}f^{2}\left(  \beta\mathcal{X}\right)  =\frac{E^{2}}{h\left(
r\right)  }-m^{2},\label{eq:S2}%
\end{equation}
where we have%
\begin{equation}
\mathcal{X=}h\left(  r\right)  \left(  \frac{dS_{1}}{dr}\right)  ^{2}%
+\frac{L^{2}}{r^{2}}.
\end{equation}
Solving eqn. $\left(  \ref{eq:S2}\right)  $ gives%
\begin{equation}
S_{1}=\int dr\sqrt{\frac{1}{h\left(  r\right)  }\left(  \frac{E^{2}}{h\left(
r\right)  }-m^{2}\right)  g\left(  \beta\left(  \frac{E^{2}}{h\left(
r\right)  }-m^{2}\right)  \right)  -\frac{L^{2}}{h\left(  r\right)  r^{2}}},
\end{equation}
where $g\left(  x\right)  =1-2x+\mathcal{O}\left(  x^{2}\right)  $ for the
Brau reduction. Because of the spherical symmetry of the Schwarzschild metric
we can therefore, with no loss of generality, confine our attention to
particles moving in the equatorial plane given by $\theta=\frac{\pi}{2}$. In
this case, one has that $\sin\theta=1$, $L=L_{z}$, and the trajectory is
\begin{equation}
\phi=-\frac{\partial S_{1}}{\partial L}=L\int\frac{dr}{r^{2}\sqrt{\left[
E^{2}-h\left(  r\right)  m^{2}\right]  g\left(  \beta\left(  \frac{E^{2}%
}{h\left(  r\right)  }-m^{2}\right)  \right)  -\frac{h\left(  r\right)  L^{2}%
}{r^{2}}}}.\label{eq:phiRC}%
\end{equation}
The time-dependence of the motion is then obtained by the inverse Legendre
transformation:%
\begin{equation}
t=\frac{\partial S_{1}}{\partial E}.\label{eq:time}%
\end{equation}

\subsubsection{Precession of Planetary Orbits}

For massive particles, differentiate eqn. $\left(  \ref{eq:phiRC}\right)  $
with respect to $r$ gives%
\begin{equation}
\left(  \frac{Adu}{d\phi}\right)  ^{2}=\left[  \tilde{E}^{2}-\left(
1+\varepsilon\right)  \left(  1-2uA^{2}\right)  \right]  -\left(
1-2uA^{2}\right)  A^{2}u^{2}-\varepsilon\frac{\tilde{E}^{4}}{\left(
1-2uA^{2}\right)  }+2\varepsilon\tilde{E}^{2}+\mathcal{O}\left(
\varepsilon^{2}\right)  , \label{eq:u}%
\end{equation}
where we have%
\begin{equation}
A=\frac{GMm}{L}\text{, }u=\frac{L^{2}}{GMm^{2}r}\text{, }\tilde{E}=\frac{E}%
{m}\text{, and }\varepsilon=2\beta m^{2}.
\end{equation}
One then differentiates eqn. $\left(  \ref{eq:u}\right)  $ with respect to
$\phi$ and obtains a second-order equation for $u\left(  \phi\right)  $:%
\begin{equation}
\frac{d^{2}u}{d\phi^{2}}=1+\varepsilon\left(  1-\tilde{E}^{4}\right)
-u-4\varepsilon\tilde{E}^{4}A^{2}u+3A^{2}u^{2}+\mathcal{O}\left(
A^{4},\varepsilon^{2}\right)  . \label{eq:uandphi}%
\end{equation}
For $A$, $\varepsilon\ll1$, we put $u$ in a form of a Taylor series in terms
of $\varepsilon$ and $A^{2}$:
\begin{equation}
u=u_{0}+A^{2}x+\varepsilon y+A^{2}\varepsilon z+\mathcal{O}\left(
A^{4},\varepsilon^{2}\right)  , \label{eq:uzero}%
\end{equation}
where $u_{0}$ is a Newtonian solution while $A^{2}x$ is a small deviation due
to general relativity, and $\varepsilon y$ and $A^{2}\varepsilon z$ due to
quantum gravity. Plugging eqn. $\left(  \ref{eq:uzero}\right)  \,$into eqn.
$\left(  \ref{eq:uandphi}\right)  $, we obtain
\begin{align}
\frac{d^{2}u_{0}}{d\phi^{2}}+u_{0}  &  =1,\nonumber\\
\frac{d^{2}x}{d\phi^{2}}+x  &  =3u_{0}^{2},\label{eq:xyz}\\
\frac{d^{2}y}{d\phi^{2}}+y  &  =\left(  1-\tilde{E}^{4}\right)  ,\nonumber\\
\frac{d^{2}z}{d\phi^{2}}+z  &  =-4\tilde{E}^{4}u_{0}+6yu_{0}.\nonumber
\end{align}

For a bound orbit of a planet, the first equation in eqns. $\left(
\ref{eq:xyz}\right)  $ has the solution:%
\begin{equation}
u_{0}=1+e\cos\phi\text{,} \label{eq:u0}%
\end{equation}
which describes an ellipse with the eccentricity $e$. It follows from eqn.
$\left(  \ref{eq:u0}\right)  $ that the rest equations of eqns. $\left(
\ref{eq:xyz}\right)  $ give%
\begin{align}
x  &  =3\left(  1+\frac{1}{2}e^{2}\right)  +3e\phi\sin\phi-\frac{1}{2}%
e^{2}\cos2\phi,\nonumber\\
y  &  =\left(  1-\tilde{E}^{4}\right)  ,\label{eq:solutionxyz}\\
z  &  =\left(  6-10\tilde{E}^{4}\right)  +\left(  3-5\tilde{E}^{4}\right)
e\phi\sin\phi.\nonumber
\end{align}
The first terms in expressions of $x$, $y$, and $z$ in eqns. $\left(
\ref{eq:solutionxyz}\right)  $ are constant displacement, while the last ones
in expressions of $x$ and $z$ oscillate around zero. However, the terms with
$\phi\sin\phi$ describe effects which accumulate over successive orbits.
Combing these terms with $u_{0},$ we have
\begin{equation}
u=1+e\cos\left[  \left(  1-\alpha\right)  \phi\right]  ,
\end{equation}
where%
\begin{equation}
\alpha=3A^{2}\left[  1+\varepsilon\left(  1-\frac{5\tilde{E}^{4}}{3}\right)
\right]  +\mathcal{O}\left(  A^{4},\varepsilon^{2}\right)  .
\end{equation}
We find, during each orbit of the planet, perihelion advances by an angle:%
\begin{equation}
\Delta\omega=2\pi\alpha,
\end{equation}
and the contribution from the minimal length is
\begin{equation}
\Delta\omega_{\beta}=2\pi\left(  3A^{2}\right)  \varepsilon\left(
1-\frac{5\tilde{E}^{4}}{3}\right)  +\mathcal{O}\left(  \varepsilon^{2}\right)
. \label{eq:ourcorrection}%
\end{equation}
Since $E$ is the energy of a planet including its rest energy, one has
$\tilde{E}^{2}=1+\mathcal{O}\left(  A^{2}\right)  $ and hence
\begin{equation}
\Delta\omega_{\beta}\approx-2\pi\frac{4\beta GMm^{2}}{a\left(  1-e^{2}\right)
}, \label{eq:prosession-GR}%
\end{equation}
where we use $L^{2}=GMm^{2}\left(  1-e^{2}\right)  a.$ It follows from eqn.
$\left(  \ref{eq:prosession-GR}\right)  $ that $\Delta\omega_{\beta}$ obtained
in the context of deformed general relativity is the same as $\Delta
\omega_{\beta}$ in eqn. $\left(  \ref{eq:procession-NM}\right)  $\ and
\cite{IN-Benczik:2002tt} with $\beta^{\prime}=2\beta$, which have been
computed in the context of deformed Newtonian dynamics.

\subsubsection{Deflection of Light}

Since a massive object can have a significant effect on the propagation of
photons, we can test the predictions of general relativity by investigating
the slight deflection of light by, for example, the Sun.

For massless particles, eqn. $\left(  \ref{eq:phiRC}\right)  $ leads to a
second-order equation for $u\left(  \phi\right)  $:%
\begin{equation}
\frac{d^{2}u}{d\phi^{2}}=-\varepsilon\left(  1+4A^{2}u\right)  -u+3A^{2}u^{2},
\label{eq:uandphimassless}%
\end{equation}
where
\begin{equation}
u=\frac{L^{2}}{GME^{2}r}\text{, }\varepsilon=2\beta E^{2}\text{, and }%
A=\frac{GME}{L}\text{.}%
\end{equation}
Plugging the ansatz $\left(  \ref{eq:uzero}\right)  \,$into eqn. $\left(
\ref{eq:uandphimassless}\right)  $, we obtain
\begin{align}
\frac{d^{2}u_{0}}{d\phi^{2}}+u_{0}  &  =0,\nonumber\\
\frac{d^{2}x}{d\phi^{2}}+x  &  =3u_{0}^{2},\label{eq:xyzmassless}\\
\frac{d^{2}y}{d\phi^{2}}+y  &  =-1,\nonumber\\
\frac{d^{2}z}{d\phi^{2}}+z  &  =-4u_{0}+6yu_{0}.\nonumber
\end{align}
In the absence of matter, we may write the solution for $u_{0}$ as%
\begin{equation}
u_{0}=B\sin\phi, \label{eq:u0massless}%
\end{equation}
which represents a straight-line path with impact parameter $b=\frac{L^{2}%
}{GME^{2}B}$. Solving eqns. $\left(  \ref{eq:xyz}\right)  $ for $x$, $y$, and
$z$ gives%
\begin{align}
x  &  =\frac{3B^{2}}{2}\left(  1+\frac{1}{3}\cos2\phi\right)  ,\nonumber\\
y  &  =-1,\label{eq:solutionxyzmassless}\\
z  &  =5B\phi\cos\phi.\nonumber
\end{align}
Combining eqn. $\left(  \ref{eq:u0massless}\right)  $ with eqns. $\left(
\ref{eq:solutionxyzmassless}\right)  $, we find%
\begin{equation}
u=B\sin\left[  \left(  1-\alpha\right)  \phi\right]  +\frac{3A^{2}B^{2}}%
{2}\left(  1+\frac{1}{3}\cos\left[  2\left(  1-\alpha\right)  \phi\right]
\right)  -\varepsilon+\mathcal{O}\left(  A^{4},\varepsilon^{2}\right)  ,
\end{equation}
where%
\begin{equation}
\alpha=-5\varepsilon A^{2}.
\end{equation}

Now consider the limit $r\rightarrow\infty$, i.e. $u\rightarrow0$. Clearly,
for a slight deflection we can take $\sin\phi\approx\phi$ and cos$\phi
\approx1$ at infinity, to obtain%
\begin{equation}
\phi=-2A^{2}B\left(  1-\frac{\varepsilon}{2A^{2}B^{2}}-5\varepsilon
A^{2}\right)  +\mathcal{O}\left(  A^{4},\varepsilon^{2}\right)  .
\end{equation}
Thus the total deflection is%
\begin{equation}
\Delta\phi=\frac{4GM}{b}\left(  1-\frac{\varepsilon}{2}-5\varepsilon
A^{2}\right)  +\mathcal{O}\left(  A^{4},\varepsilon^{2}\right)  ,
\label{eq:deflection}%
\end{equation}
where we use $L=Eb$ for the leading order.

\subsubsection{Time-delay in Radar Propagation}

Now consider the trajectory of a photon from the observer to the test object.
Obviously, the trajectory will be deflected when the photon passes through the
gravitational field of a massive object of mass $M$. The time taken to travel
between the observer and test object can be given by eqn. $\left(
\ref{eq:time}\right)  $ with $m=0$. Let $r_{0}$ be the coordinate distance of
closest approach of the photon to the massive object. Thus, we have%
\begin{equation}
\left(  \frac{dr}{dt}\right)  _{r_{0}}=0.
\end{equation}
It then follows from eqn. $\left(  \ref{eq:time}\right)  $ that%
\begin{equation}
\frac{E^{2}}{L^{2}}=\frac{h\left(  r_{0}\right)  }{r_{0}^{2}g\left(
\frac{\beta E^{2}}{h\left(  r_{0}\right)  }\right)  }. \label{eq:r0}%
\end{equation}
Using eqns. $\left(  \ref{eq:time}\right)  $ and $\left(  \ref{eq:r0}\right)
$, we find that the time taken to travel between points $r_{0}$ and $r$ is%
\begin{equation}
t\left(  r,r_{0}\right)  =\int_{r_{0}}^{r}dr\frac{g\left(  \frac{\varepsilon
}{h\left(  r\right)  }\right)  +g^{\prime}\left(  \frac{\varepsilon}{h\left(
r\right)  }\right)  \frac{\varepsilon}{h\left(  r\right)  }}{\sqrt{h\left(
r_{0}\right)  r^{2}g\left(  \frac{\varepsilon}{h\left(  r\right)  }\right)
-h\left(  r\right)  r_{0}^{2}g\left(  \frac{\varepsilon}{h\left(
r_{0}\right)  }\right)  }}\frac{r\sqrt{h\left(  r_{0}\right)  }}{h\left(
r\right)  },
\end{equation}
where $\varepsilon=$ $2\beta E^{2}$. Integrating this leads to%
\begin{align}
t\left(  r,r_{0}\right)   &  =\left(  1-\frac{3\varepsilon}{2}\right)
\sqrt{r^{2}-r_{0}^{2}}+GM\sqrt{\frac{r-r_{0}}{r+r_{0}}}\left(  1-\frac
{5\varepsilon}{2}\right) \nonumber\\
&  +2\left(  1-3\varepsilon\right)  GM\ln\left(  \frac{r+\sqrt{r^{2}-r_{0}%
^{2}}}{r_{0}}\right)  +\mathcal{O}\left(  \varepsilon^{2},\frac{G^{2}M^{2}%
}{r_{0}^{2}}\right)  . \label{eq:tandt0}%
\end{align}
Note that eqn. $\left(  \ref{eq:HJFD}\right)  $ gives the energy-momentum
dispersion relation of a photon in flat spacetime:%
\begin{equation}
E\approx p\left(  1+\beta p^{2}\right)  ,
\end{equation}
where we use $E=\frac{\partial S}{\partial t}$, $p_{i}=\frac{\partial
S}{\partial x_{i}}$, and $p^{2}=p^{i}p_{i}$. Thus, after effects of the
minimal length are considered, the light speed in flat spacetime becomes
\begin{equation}
\frac{\partial E}{\partial p}\approx1+3\beta p^{2}\approx1+\frac{3\varepsilon
}{2},
\end{equation}
which gives that the first term in eqn. $\left(  \ref{eq:tandt0}\right)  $ is
just what we would have expected if the light had been travelling in flat
spacetime along a straight line. The second and third terms give us the extra
time taken for the photon to travel along the curved path. If a radar beam is
sent to the test object and bounces back to the observer, the excess time
delay over a straight-line path is%
\begin{equation}
\Delta t=2\left[  t\left(  r_{1},r_{0}\right)  +t\left(  r_{2},r_{0}\right)
-t\left(  r_{1},r_{0}\right)  |_{G=0}-t\left(  r_{2},r_{0}\right)
|_{G=0}\right]  ,
\end{equation}
where $r_{1}$ and $r_{2}$ (both assumed $\gg r_{0}$) are the distances of the
observer and test object from the massive object, respectively. Thus one
obtains%
\begin{equation}
t\left(  r_{1},r_{0}\right)  -t\left(  r_{1},r_{0}\right)  |_{G=0}=GM\left(
1-\frac{5\varepsilon}{2}\right)  +2\left(  1-3\varepsilon\right)  GM\ln\left(
\frac{r_{1}}{r_{0}}\right)  ,
\end{equation}
and likewise for $t\left(  r_{2},r_{0}\right)  $. In this case, the time delay
becomes%
\begin{equation}
\Delta t\approx4GM\left[  \left(  1-\frac{5\varepsilon}{2}\right)  +\left(
1-3\varepsilon\right)  \ln\left(  \frac{r_{1}r_{2}}{r_{0}^{2}}\right)
\right]  . \label{eq:timedelay}%
\end{equation}

\section{Comparison with Experiments}

\label{Sec:CE}

To make comparison with experiments, we often express the parameter $\beta$ in
terms of a dimensionless parameter $\beta_{0}$:%
\begin{equation}
\beta=\beta_{0}\ell_{p}^{2}/\hbar^{2}=\beta_{0}/m_{pl}^{2},
\end{equation}
where $m_{pl}$ is the Planck mass, and $\ell_{p}$ is the Planck length. For
the Brau reduction, the minimal length associated with $\beta$ is $\Delta
X_{\min}=\hbar\sqrt{3\beta}=\sqrt{3\beta_{0}}\ell_{p}$. Naturally, if the
minimal length is assumed to be order of the Planck length $\ell_{p}$, one has
$\beta_{0}\sim1$. In \cite{CE-Das:2008kaa}, based on discussions of effects of
the minimal length on the tunnelling current in a Scanning Tunnelling
Microscope, an upper bound of $\beta_{0}$ was given by $\beta_{0}<10^{21}$.

We calculate the precession angle of a planet caused by deformations in
context of Newtonian dynamics in section \ref{Sec:NR} and general relativity
in section \ref{Sec:RC}. Both of our results $\left(  \ref{eq:procession-NM}%
\right)  $ and $\left(  \ref{eq:prosession-GR}\right)  $ perfectly coincide
with the result of \cite{IN-Benczik:2002tt} (eqn. $\left(  66\right)  $) with
$\beta^{\prime}=2\beta$. In \cite{IN-Benczik:2002tt}, the authors compared
their result to the observed precession of the perihelion of Mercury and
estimated an upper bound on $\beta$:%
\begin{equation}
\hbar\sqrt{\beta_{M}}<2.3\times10^{-68}%
\operatorname{m}%
\sim10^{-33}\ell_{p},
\end{equation}
where the subscript $M$ of $\beta$ means that $\beta$ is for Mercury. It is
quite surprising to note that this minimal length is $33$ orders of magnitude
below the Planck length. However, as pointed out in \cite{CE-Quesne:2009vc},
this strangely small result stemmed form the assumption made in
\cite{IN-Benczik:2002tt} that the deformation parameter $\beta$ for Mercury
was the same as for elementary particles. It also was shown in
\cite{CE-Quesne:2009vc} that if Mercury consists of $N$ quarks, the
deformation parameter $\beta_{M}$ was substantially reduced by a factor
$N^{-2}$:%
\begin{equation}
\beta_{M}=\frac{\beta_{q}}{N^{2}},
\end{equation}
where $\beta_{q}$ is $\beta$ for quarks. Since $N\sim10^{50}$, the upper
bounds on the deformation parameter $\beta$ for quarks was given by%
\begin{equation}
\hbar\sqrt{\beta_{q}}<1.4\times10^{-17}%
\operatorname{m}%
\sim10^{18}\ell_{p}\text{ and }\beta_{0}^{q}<10^{36}.
\end{equation}

For the observational tests of general relativity involving null geodesics, we
calculate the spatial deflection of star light by the Sun in section
\ref{Sec:RC}. From eqn. $\left(  \ref{eq:deflection}\right)  $, it follows
that the deflection angle of a photon's trajectory caused by deformations is
\begin{equation}
\Delta\phi_{\beta}\equiv\Delta\phi-\Delta\phi_{0}\approx-\Delta\phi_{0}%
\beta_{p}E^{2},
\end{equation}
where the third term in the bracket of eqn. $\left(  \ref{eq:deflection}%
\right)  $ is neglected since $A\ll1$, $\Delta\phi_{0}\equiv\frac{4GM}{b}$ is
the deflection angle calculated in context of general relativity without
deformation, $E$ is the energy of photons, and $\beta_{p}$ is the deformation
parameter $\beta$ for photons. For light grazing the Sun it yields
\begin{equation}
\Delta\phi_{0}=1.75%
\operatorname{{}^{\prime \prime}}%
. \label{eq:deltaphizero}%
\end{equation}
Since the 1919 eclipse expedition led by Eddington to measure the deflection
angle, several similar experiments were conducted. The Texas expedition to
Chinguetti Oasis, Mauritania, at the eclipse of 30 June 1973 gave
\cite{CE-Brune:1976jc}%
\begin{equation}
\Delta\phi_{\text{obs}}=\left(  0.95\pm0.11\right)  \Delta\phi_{0},
\label{eq:deltaphi}%
\end{equation}
where the error was $1\sigma$. Comparison of eqns. $\left(
\ref{eq:deltaphizero}\right)  $ and $\left(  \ref{eq:deltaphi}\right)  $
places a lower bound on $\Delta\phi_{\beta}$:%
\begin{equation}
\Delta\phi_{\text{obs}}-\Delta\phi_{0}=\left(  -0.05\pm0.11\right)  \Delta
\phi_{0}<\Delta\phi_{\beta}. \label{eq:deltaphisub}%
\end{equation}
At $3\sigma$ eqn. $\left(  \ref{eq:deltaphisub}\right)  $ gives%
\begin{equation}
\beta_{0}^{p}<10^{55}, \label{eq:betaq}%
\end{equation}
where $E=\frac{2\pi\hbar c}{\lambda}$, and we assume that $\lambda\sim500%
\operatorname{nm}%
$ for visible light. On the other hand, the tightest observational constraint
to date on $\Delta\phi$ comes from observations of $87$ VLBI sites and $541$
radio sources over a period of 20 years. The typical frequencies of radio
sources are around $10%
\operatorname{GHz}%
$ \cite{CE-Lebach:1995zz}. The result of this is \cite{CE-Shapiro:2004zz}%
\begin{equation}
\Delta\phi_{\text{obs}}=\left(  0.99992\pm0.00023\right)  \Delta\phi_{0},
\end{equation}
which is around $3$ orders of magnitude better than the observations of
eclipse expeditions. Similarly, one has at $3\sigma$ that%
\begin{equation}
\beta_{0}^{p}<10^{62},
\end{equation}
which is much less stringent than eqn. $\left(  \ref{eq:betaq}\right)  $ since
the typical energies of radio sources \cite{CE-Shapiro:2004zz} are much less
than these of visible light observed in eclipse expeditions.

A currently more constraining test of general relativity using null
trajectories involves the Shapiro time-delay effect which has been studied in
section \ref{Sec:RC}. From eqn. $\left(  \ref{eq:timedelay}\right)  $, it
follows that time delay over a straight-line path caused by deformations is%
\begin{equation}
\Delta t_{\beta}\equiv\Delta t-\Delta t_{0}=-\frac{\varepsilon\Delta t_{0}}%
{2}\frac{5+6\ln\left(  \frac{r_{1}r_{2}}{r_{0}^{2}}\right)  }{1+\ln\left(
\frac{r_{1}r_{2}}{r_{0}^{2}}\right)  }\approx-3\varepsilon\Delta t_{0},
\end{equation}
where we use $r_{1}$, $r_{2}\gg r_{0}$, and $\Delta t_{0}=4GM\left[
1+\ln\left(  \frac{r_{1}r_{2}}{r_{0}^{2}}\right)  \right]  $ is what we expect
in the context of general relativity without deformation. In the gravitational
field of the sun, the best constraint on this time-delay effect is obtained by
using radio links with the Cassini spacecraft between the 6th of June and the
7th of July 2002 \cite{CE-Bertotti:2003rm}. These observations result in the
constraint%
\begin{equation}
\Delta t_{\text{obs}}=\left(  1.00001\pm0.00001\right)  \Delta t_{0}.
\end{equation}
The typical frequencies of radio photons transmitted from the ground to the
Cassini spacecraft are around $10$MHz. At $3\sigma$ one has an upper bound on
$\beta_{0}^{p}$:%
\begin{equation}
\beta_{0}^{p}<10^{66}\text{.}%
\end{equation}

\section{Discussion and Conclusion}

\label{Sec:Con}

In this paper, we have used the Hamilton-Jacobi method to investigate effects
of the minimal length on the classical orbits of particles in a gravitation
field. Specifically, we derived the deformed Hamilton-Jacobi equation and used
it to study the precession of planetary orbits in the context of deformed
Newtonian dynamics. In the context of deformed general relativity, the
deformed Hamilton-Jacobi equation in the Schwarzschild metric has also been
obtained to calculate the precession angle of planetary orbits, deflection
angle of light, and time delay in radar propagation. Comparison with the
observational results places constraints on the deformation parameter
$\beta_{0}$.

In \cite{IN-Benczik:2002tt}, the precession of planetary orbits has also been
studied in the classical limit of deformed spaces using the deformed Poisson
bracket. The precession angle caused by deformations was calculated in the
context of deformed Newtonian dynamics. Our calculations confirm their results
not only in the context of deformed Newtonian dynamics but also in the context
of deformed general relativity, at least to the leading order in $\beta$.

In \cite{IN-Scardigli:2014qka}, the authors introduced the deformed
Schwarzschild metric to reproduce the Hawking temperature derived from the
deformed fundamental commutation relation $\left(  \ref{eq:1dGUP}\right)  $.
Using this deformed metric, they computed corrections to the standard general
Relativistic predictions for the light deflection and perihelion precession.
Specifically, the deformed Schwarzschild metric takes the form:%
\begin{equation}
ds^{2}=F\left(  r\right)  dt^{2}-F^{-1}\left(  r\right)  dr^{2}-r^{2}%
d\Omega^{2}, \label{eq:deformedSchMetric}%
\end{equation}
where $F\left(  r\right)  =1-\frac{R_{H}}{r}+\tilde{\varepsilon}\frac
{R_{H}^{n}}{2^{n}r^{n}}$, $R_{H}=2GM$, and the case with $n=2$ was considered
in \cite{IN-Scardigli:2014qka}. Following calculations in
\cite{IN-Scardigli:2014qka}, one finds that the horizons of the metric
$\left(  \ref{eq:deformedSchMetric}\right)  $ is%
\[
r_{H}=R_{H}\left(  1-2^{-n}\tilde{\varepsilon}\right)  +\mathcal{O}\left(
\tilde{\varepsilon}^{2}\right)  ,
\]
and the Hawking temperature is%
\begin{equation}
T\left(  \varepsilon\right)  =\frac{\hbar}{4\pi R_{H}}\left[  1+2^{-n}\left(
2-n\right)  \tilde{\varepsilon}\right]  +\mathcal{O}\left(  \tilde
{\varepsilon}^{2}\right)  .
\end{equation}
Using eqn. $\left(  2.31\right)  $ in \cite{IN-Scardigli:2014qka}, one could
relate $\tilde{\varepsilon}$ to $\beta_{0}$ (which is $\beta$ in
\cite{IN-Scardigli:2014qka}) in the cases with $n\neq2$:%
\begin{equation}
\tilde{\varepsilon}\approx\frac{2^{n-4}\beta_{0}\hbar}{\pi^{2}GM^{2}\left(
2-n\right)  }. \label{eq:epsilonn!=2}%
\end{equation}
For the case $n=2$, it was shown in \cite{IN-Scardigli:2014qka} that%
\begin{equation}
\tilde{\varepsilon}^{2}\approx-\frac{\hbar\beta_{0}}{\pi^{2}GM^{2}},
\label{eq:epsilon=2}%
\end{equation}
which required that $\beta_{0}<0$.

By contrast, there are a number of differences between the methods used in our
paper and in \cite{IN-Scardigli:2014qka}, which are as follows:

\begin{enumerate}
\item The authors of \cite{IN-Scardigli:2014qka} calculated the precession
angle of Mercury's orbits in the deformed Schwarzschild metric $\left(
\ref{eq:deformedSchMetric}\right)  $ with $n=2$ and found the correction was
\begin{equation}
\frac{\Delta\phi-\Delta\phi_{0}}{\Delta\phi_{0}}\sim\tilde{\varepsilon},
\label{eq:correction}%
\end{equation}
which only depended on the mass of the Sun. It is naturally expected that eqn.
$\left(  \ref{eq:correction}\right)  $ is also true for the cases with
$n\neq2$. According to eqns. $\left(  \ref{eq:epsilonn!=2}\right)  $ and
$\left(  \ref{eq:epsilon=2}\right)  $, the correction $\left(
\ref{eq:correction}\right)  $ is proportional to $\sqrt{\beta_{0}}$ in the
$n=2$ case while it is proportional to $\beta_{0}$ in the $n\neq2$ cases. On
the other hand, we find that results in our paper and \cite{IN-Benczik:2002tt}
are only proportional to $\beta_{0}$. It seems that results obtained using the
method proposed in \cite{IN-Scardigli:2014qka} may depend on the ansatz form
of $F\left(  r\right)  $. It also follows from eqn. $\left(
\ref{eq:epsilonn!=2}\right)  $ that one does not need to require $\beta_{0}<0$
in the $n\neq2$ cases.

\item Moreover, the results in our paper (see eqn. $\left(
\ref{eq:ourcorrection}\right)  $) and \cite{IN-Benczik:2002tt} show that%
\begin{equation}
\frac{\Delta\phi-\Delta\phi_{0}}{\Delta\phi_{0}}\sim\varepsilon=2\beta
m^{2}\sim\beta E^{2}, \label{eq:our-correction}%
\end{equation}
where we use $E\approx m$ for Mercury. The correction $\left(
\ref{eq:our-correction}\right)  $ depends on the energy $E$ of Mercury while
the correction $\left(  \ref{eq:correction}\right)  $ obtained in
\cite{IN-Scardigli:2014qka} does not. How can we reconcile this contradiction?
One might note that the authors of \cite{IN-Scardigli:2014qka} used eqn.
$\left(  2.20\right)  $ from \cite{IN-Scardigli:2014qka}: $E=T,$ to express
the deformed Hawking temperature $T$ in terms of the mass of the Schwarzschild
black holes. If one substitutes
\begin{equation}
E=T\sim\frac{\hbar}{8\pi GM},
\end{equation}
into eqn. $\left(  \ref{eq:our-correction}\right)  $, we find
\begin{equation}
\varepsilon\sim\frac{\beta_{0}\hbar}{16\pi^{2}GM^{2}}, \label{eq:ourepslion}%
\end{equation}
where we use $\frac{\hbar}{G}=4m_{pl}^{2}$. It follows that eqn. $\left(
\ref{eq:ourepslion}\right)  $ is the same as eqns. $\left(
\ref{eq:epsilonn!=2}\right)  $ and $\left(  \ref{eq:epsilon=2}\right)  $ up to
some numerical factors. In other words, there is implicit assumption made in
\cite{IN-Scardigli:2014qka} that the energy $E$ of Mercury was given by
$E=T_{s}$, where $T_{s}$ was the Hawking temperature of the Schwarzschild
black hole of $1$ solar mass. Since $T_{s}$ is far less than the mass of
Mercury, one could expect that the results in our paper and
\cite{IN-Benczik:2002tt} place a much stronger constraint on $\beta$. In fact,
$\beta_{0}^{M}<10^{-66}$ was given in \cite{IN-Benczik:2002tt} while
$\beta_{0}^{M}<10^{72}$ in \cite{IN-Scardigli:2014qka}. To deal with the
energy $E$ of Mercury in a more appropriate way using the method proposed in
\cite{IN-Scardigli:2014qka}, one might need to resort to Gravity's rainbow
\cite{CON-Magueijo:2002xx}, where the minimal length deformations to the
Schwarzschild black hole could depend on the energy of Mercury. This is
expected since GUP is closely related to Doubly Special Relativity and
Gravity's rainbow \cite{CON-Hossenfelder:2012jw}. Another way to understand
this is to note that the deformed Hawking temperatures obtained using the
Hamilton-Jacobi method \cite{IN-Chen:2013tha,IN-Chen:2013ssa,IN-Chen:2014xgj}%
\ do depend on the energy of radiated particles. In this case, one possible
way to find the deformations to the rainbow metric is using the deformed
Hawking temperatures obtained in
\cite{IN-Chen:2013tha,IN-Chen:2013ssa,IN-Chen:2014xgj} instead.
\end{enumerate}

Finally, we used the observational results to places constraints on $\beta
_{0}$ in section \ref{Sec:CE}. Comparing with constraints on $\beta_{0}$ from
other papers, our results are much less stringent. In other words, it is
difficult to observe quantum gravity effects on the deformed classical motions
of particles. One of reasons for these difficulties is that the energy of the
particles in classical motions is too small compared to the Planck mass
$m_{pl}$. Typically, the corrections due to the minimal length is around
$\beta_{0}E^{2}/m_{pl}^{2}$. For photons, one has $E\sim1%
\operatorname{eV}%
$ for visible light and $E\sim10^{-4}%
\operatorname{eV}%
$ for radio of frequency $10%
\operatorname{GHz}%
$. This also explains why the observations of eclipse expeditions put a
stronger constraint on $\beta_{0}$ even though observations of VLBI have $3$
orders of magnitude better precision. For nonrelativistic massive particles of
mass $m$ in a weak gravitational field, one has $E\sim m$. Thus, the minimal
length correction to the precession angle of planetary orbits is around
$\frac{\beta_{0}^{q}m_{\text{nul}}^{2}}{3^{2}m_{pl}^{2}}$, where
$m_{\text{nul}}\sim1%
\operatorname{GeV}%
$ is the mass of a nucleon. It follows that the observations of the precession
of Mercury would place the strongest constraint on $\beta_{0}$ in our paper.

\begin{acknowledgments}
We are grateful to Houwen Wu and Zheng Sun for useful discussions. This work
is supported in part by NSFC (Grant No. 11005016, 11175039 and 11375121),
Doctoral Fund of Southwest University of Science and Technology (Grant No.
10zx7139), and the Fundamental Research Funds for the Central Universities.
\end{acknowledgments}


\noindent

\begin{thebibliography}{99}                                                                                               %


\bibitem {IN-Veneziano:1986zf}G.~Veneziano, ``A Stringy Nature Needs Just Two
Constants,'' Europhys.\ Lett.\ \textbf{2}, 199 (1986). doi:10.1209/0295-5075/2/3/006

\bibitem {IN-Gross:1987ar}D.~J.~Gross and P.~F.~Mende, ``String Theory Beyond
the Planck Scale,'' Nucl.\ Phys.\ B \textbf{303}, 407 (1988). doi:10.1016/0550-3213(88)90390-2

\bibitem {IN-Amati:1988tn}D.~Amati, M.~Ciafaloni and G.~Veneziano, ``Can
Space-Time Be Probed Below the String Size?,'' Phys.\ Lett.\ B \textbf{216},
41 (1989). doi:10.1016/0370-2693(89)91366-X

\bibitem {IN-Garay:1994en}L.~J.~Garay, ``Quantum gravity and minimum length,''
Int.\ J.\ Mod.\ Phys.\ A \textbf{10}, 145 (1995) doi:10.1142/S0217751X95000085 [gr-qc/9403008].

\bibitem {IN-Maggiore:1993kv}M.~Maggiore, ``The Algebraic structure of the
generalized uncertainty principle,'' Phys.\ Lett.\ B \textbf{319}, 83 (1993)
doi:10.1016/0370-2693(93)90785-G [hep-th/9309034].

\bibitem {IN-Kempf:1994su}A.~Kempf, G.~Mangano and R.~B.~Mann, ``Hilbert space
representation of the minimal length uncertainty relation,'' Phys.\ Rev.\ D
\textbf{52}, 1108 (1995) doi:10.1103/PhysRevD.52.1108 [hep-th/9412167].

\bibitem {IN-Hossenfelder:2012jw}S.~Hossenfelder, \textquotedblleft Minimal
Length Scale Scenarios for Quantum Gravity,\textquotedblright\ Living
Rev.\ Rel.\ \textbf{16}, 2 (2013) doi:10.12942/lrr-2013-2 [arXiv:1203.6191 [gr-qc]].

\bibitem {IN-Chang:2001kn}L.~N.~Chang, D.~Minic, N.~Okamura and T.~Takeuchi,
``Exact solution of the harmonic oscillator in arbitrary dimensions with
minimal length uncertainty relations,'' Phys.\ Rev.\ D \textbf{65}, 125027
(2002) doi:10.1103/PhysRevD.65.125027 [hep-th/0111181].

\bibitem {IN-Akhoury:2003kc}R.~Akhoury and Y.~P.~Yao, ``Minimal length
uncertainty relation and the hydrogen spectrum,'' Phys.\ Lett.\ B
\textbf{572}, 37 (2003) doi:10.1016/j.physletb.2003.07.084 [hep-ph/0302108].

\bibitem {IN-Brau:1999uv}F.~Brau, \textquotedblleft Minimal length uncertainty
relation and hydrogen atom,\textquotedblright\ J.\ Phys.\ A \textbf{32}, 7691
(1999) doi10.1088/0305-4470/32/44/308 [quant-ph/9905033].

\bibitem {IN-Brau:2006ca}F.~Brau and F.~Buisseret, ``Minimal Length
Uncertainty Relation and gravitational quantum well,'' Phys.\ Rev.\ D
\textbf{74}, 036002 (2006) doi:10.1103/PhysRevD.74.036002 [hep-th/0605183].

\bibitem {IN-Pedram:2011xj}P.~Pedram, K.~Nozari and S.~H.~Taheri,
\textquotedblleft The effects of minimal length and maximal momentum on the
transition rate of ultra cold neutrons in gravitational
field,\textquotedblright\ JHEP \textbf{1103}, 093 (2011)
doi:10.1007/JHEP03(2011)093 [arXiv:1103.1015 [hep-th]].

\bibitem {IN-Benczik:2002tt}S.~Benczik, L.~N.~Chang, D.~Minic, N.~Okamura,
S.~Rayyan and T.~Takeuchi, \textquotedblleft Short distance versus long
distance physicsThe Classical limit of the minimal length uncertainty
relation,\textquotedblright\ Phys.\ Rev.\ D \textbf{66}, 026003 (2002)
doi10.1103/PhysRevD.66.026003 [hep-th/0204049].

\bibitem {IN-Jalalzadeh:2014jea}S.~Jalalzadeh, S.~M.~M.~Rasouli and
P.~V.~Moniz, \textquotedblleft Quantum cosmology, minimal length and
holography,\textquotedblright\ Phys.\ Rev.\ D \textbf{90}, no. 2, 023541
(2014) doi:10.1103/PhysRevD.90.023541 [arXiv:1403.1419 [gr-qc]].

\bibitem {IN-Quintela:2015bua}T.~S.~Quintela, Jr., J.~C.~Fabris and
J.~A.~Nogueira, ``The Harmonic Oscillator in the Classical Limit of a
Minimal-Length Scenario,'' arXiv:1510.08129 [hep-th].

\bibitem {IN-Tkachuk:2013qa}V.~M.~Tkachuk, \textquotedblleft Deformed
Heisenberg algebra with minimal length and equivalence
principle,\textquotedblright\ Phys.\ Rev.\ A \textbf{86}, 062112 (2012)
doi:10.1103/PhysRevA.86.062112 [arXiv:1301.1891 [gr-qc]].

\bibitem {IN-Silagadze:2009vu}Z.~K.~Silagadze, \textquotedblleft Quantum
gravity, minimum length and Keplerian orbits,\textquotedblright%
\ Phys.\ Lett.\ A \textbf{373}, 2643 (2009) doi:10.1016/j.physleta.2009.05.053
[arXiv:0901.1258 [gr-qc]].

\bibitem {IN-Ahmadi:2014cga}F.~Ahmadi and J.~Khodagholizadeh,
\textquotedblleft Effect of GUP on the Kepler problem and a variable minimal
length,\textquotedblright\ Can.\ J.\ Phys.\ \textbf{92}, 484 (2014)
doi:10.1139/cjp-2013-0354 [arXiv:1411.0241 [hep-th]].

\bibitem {IN-Scardigli:2014qka}F.~Scardigli and R.~Casadio, ``Gravitational
tests of the Generalized Uncertainty Principle,'' Eur.\ Phys.\ J.\ C
\textbf{75}, no. 9, 425 (2015) doi:10.1140/epjc/s10052-015-3635-y
[arXiv:1407.0113 [hep-th]].

\bibitem {IN-Ali:2015zua}A.~Farag Ali, M.~M.~Khalil and E.~C.~Vagenas,
\textquotedblleft Minimal Length in quantum gravity and gravitational
measurements,\textquotedblright\ Europhys.\ Lett.\ \textbf{112}, no. 2, 20005
(2015) doi:10.1209/0295-5075/112/20005 [arXiv:1510.06365 [gr-qc]].

\bibitem {IN-Tao:2012fp}J.~Tao, P.~Wang and H.~Yang, ``Homogeneous Field and
WKB Approximation In Deformed Quantum Mechanics with Minimal Length,''
Adv.\ High Energy Phys.\ \textbf{2015}, 718359 (2015) doi:10.1155/2015/718359
[arXiv:1211.5650 [hep-th]].

\bibitem {IN-Chen:2013tha}D.~Chen, H.~Wu and H.~Yang, ``Observing remnants by
fermions' tunneling,'' JCAP \textbf{1403}, 036 (2014)
doi:10.1088/1475-7516/2014/03/036 [arXiv:1307.0172 [gr-qc]].

\bibitem {IN-Chen:2013ssa}D.~Y.~Chen, Q.~Q.~Jiang, P.~Wang and H.~Yang,
``Remnants, fermions` tunnelling and effects of quantum gravity,'' JHEP
\textbf{1311}, 176 (2013) doi:10.1007/JHEP11(2013)176 [arXiv:1312.3781 [hep-th]].

\bibitem {IN-Chen:2014xgj}D.~Chen, H.~Wu, H.~Yang and S.~Yang, ``Effects of
quantum gravity on black holes,'' Int.\ J.\ Mod.\ Phys.\ A \textbf{29}, no.
26, 1430054 (2014) doi:10.1142/S0217751X14300543 [arXiv:1410.5071 [gr-qc]].

\bibitem {NC-Goldstein}H. Goldstein, \textquotedblleft Classical
Mechanics,\textquotedblright\ Addison-Wesley Press, Reading, Mass, USA, 1951.

\bibitem {RC-Kempf:1996nk}A.~Kempf and G.~Mangano, \textquotedblleft Minimal
length uncertainty relation and ultraviolet regularization,\textquotedblright%
\ Phys.\ Rev.\ D \textbf{55}, 7909 (1997) doi:10.1103/PhysRevD.55.7909 [hep-th/9612084].

\bibitem {RC-Berger:2010pj}M.~S.~Berger and M.~Maziashvili, \textquotedblleft
Free particle wavefunction in light of the minimum-length deformed quantum
mechanics and some of its phenomenological implications,\textquotedblright%
\ Phys.\ Rev.\ D \textbf{84}, 044043 (2011) doi:10.1103/PhysRevD.84.044043
[arXiv:1010.2873 [gr-qc]].

\bibitem {RC-Benrong:2014woa}M.~Benrong, P.~Wang and H.~Yang,
\textquotedblleft Deformed Hamilton-Jacobi Method in Covariant Quantum Gravity
Effective Models,\textquotedblright\ arXiv1408.5055 [gr-qc].

\bibitem {CE-Das:2008kaa}S.~Das and E.~C.~Vagenas, ``Universality of Quantum
Gravity Corrections,'' Phys.\ Rev.\ Lett.\ \textbf{101}, 221301 (2008)
doi:10.1103/PhysRevLett.101.221301 [arXiv:0810.5333 [hep-th]].

\bibitem {CE-Quesne:2009vc}C.~Quesne and V.~M.~Tkachuk, ``Composite system in
deformed space with minimal length,'' Phys.\ Rev.\ A \textbf{81}, 012106
(2010) doi:10.1103/PhysRevA.81.012106 [arXiv:0906.0050 [hep-th]].

\bibitem {CE-Brune:1976jc}R.~A.~Brune, Jr. \textit{et al.}, ``Gravitational
deflection of light: Solar eclipse of 30 June 1973 I. Description of
procedures and final results,'' Astron.\ J.\ \textbf{81}, 452 (1976). doi:10.1086/111906

\bibitem {CE-Lebach:1995zz}D.~E.~Lebach, B.~E.~Corey, I.~I.~Shapiro,
M.~I.~Ratner, J.~C.~Webber, A.~E.~E.~Rogers, J.~L.~Davis and T.~A.~Herring,
``Measurement of the Solar Gravitational Deflection of Radio Waves Using
Very-Long-Baseline Interferometry,'' Phys.\ Rev.\ Lett.\ \textbf{75}, 1439
(1995). doi:10.1103/PhysRevLett.75.1439

\bibitem {CE-Shapiro:2004zz}S.~S.~Shapiro, J.~L.~Davis, D.~E.~Lebach and
J.~S.~Gregory, ``Measurement of the Solar Gravitational Deflection of Radio
Waves using Geodetic Very-Long-Baseline Interferometry Data, 1979-1999,''
Phys.\ Rev.\ Lett.\ \textbf{92}, 121101 (2004). doi:10.1103/PhysRevLett.92.121101

\bibitem {CE-Bertotti:2003rm}B.~Bertotti, L.~Iess and P.~Tortora, ``A test of
general relativity using radio links with the Cassini spacecraft,'' Nature
\textbf{425}, 374 (2003). doi:10.1038/nature01997

\bibitem {CON-Magueijo:2002xx}J.~Magueijo and L.~Smolin, ``Gravity's
rainbow,'' Class.\ Quant.\ Grav.\ \textbf{21}, 1725 (2004)
doi:10.1088/0264-9381/21/7/001 [gr-qc/0305055].

\bibitem {CON-Hossenfelder:2012jw}S.~Hossenfelder, ``Minimal Length Scale
Scenarios for Quantum Gravity,'' Living Rev.\ Rel.\ \textbf{16}, 2 (2013)
doi:10.12942/lrr-2013-2 [arXiv:1203.6191 [gr-qc]].
\end{thebibliography}
\end{document}